\documentclass{ws-rv9x6}
\usepackage{subfigure}   
\usepackage{ws-rv-thm}   
\usepackage{ws-rv-van}   
\makeindex

\newcommand{\be}{\begin{equation}}
\newcommand{\ee}{\end{equation}}
\newcommand{\beqa}{\begin{eqnarray}}
\newcommand{\eeqa}{\end{eqnarray}}

\begin{document}

\centerline{\Large{ Noncommutative $AdS_2$  II:  The Correspondence Principle}}

\vspace{5mm}

\author[A. Stern]{Allen Stern}

\address{Department of Physics, University of Alabama,\\ Tuscaloosa,
Alabama 35487, USA, \\
astern@ua.edu}

\author[A. Stern and A. Pinzul]{Aleksandr Pinzul}

\address{Universidade  de Bras\'{\i}lia, Instituto de F\'{\i}sica\\
70910-900, Bras\'{\i}lia, DF, Brasil\\
and\\
International Center of Physics\\
C.P. 04667, Bras\'{\i}lia, DF, Brazil, \\
apinzul@unb.br}

{\it In Honor of A.P.Balachandran on the Occasion of His 85th Birthday}
\vspace{5mm}

\begin{abstract}
Using the exact solutions to the field equation for a massive scalar field on noncommutative $AdS_2$, we apply the $AdS/CFT$ correspondence principle to obtain an exact result for the associated two-point function on the conformal  boundary.  The answer satisfies conformal invariance and has the correct commutative limit and massless limit.
\end{abstract}


\body


\section{Introduction}\label{ra_sec1}
The conjectured $AdS/CFT$ correspondence principle has played a central role in theoretical physics in the past two decades.  It posits a strong/weak duality between   gravity in the bulk of an asymptotically anti-de Sitter ($AdS$) space and a conformal field theory ($CFT$)  located at  the so-called conformal boundary.\cite{Maldacena:1997re}  For obvious reasons many practical applications of the correspondence principle utilize classical, or weak, gravity in the bulk, with the intention of exploring the strong coupling regime of the boundary theory.  Since a fully consistent quantum gravity treatment of the bulk remains out of reach,  it is a nontrivial task  explore other domains of the correspondence.  On the other hand,  the incorporation of some quantum gravity effects in the bulk is possible.  This  has been the motivation for our recent works \cite{Pinzul:2017wch,deAlmeida:2019awj,Lizzi:2020cwx,Pinzul:2021cjz}.
As remarked in \cite{Stern:2022}, the inclusion of some quantum gravity effects might be achieved by replacing the $AdS$ bulk by its noncommutative analogue, $ncAdS_2$.  While, in general, the introduction of noncommutativity destroys the isometries of a manifold, and is not  unique, this is not  the case for $AdS_2$.  Therefore, $ncAdS_2$ can serve as a toy model for the introduction of quantum  gravity effects in the bulk (barring the known difficulties of the $AdS_2/CFT_1$ correspondence that already exist in the  commutative case.\footnote{ See  for example \cite{Strominger:1998yg,Maldacena:2016hyu}.}). Moreover,  in \cite{Pinzul:2017wch} it was  shown i) that the  star product for $ncAdS_2$, when acting on  functions having a well behaved boundary limit, reduces to the point-wise product in this limit,  and ii) that noncommutative corrections to isometry generators, i.e., Killing vectors, vanish near the boundary.  In other words, $ncAdS_2$ is  an asymptotically $AdS$ space and, so according to the correspondence principle, a $CFT$ should be present at its boundary.

The on-shell field theory action in the bulk $S|_{\rm on-shell}$ plays a  central role in the explicit construction of the $AdS/CFT$ correspondence.  It generates the $n-$point connected correlation functions for operators on the boundary.\cite{Witten:1998qj} As a first step, it is therefore necessary to obtain the solutions to field theories in the  bulk, which in our case is noncommutative.  Exact solutions on  $ncAdS_2$ were obtained for free massless scalar and spinor fields in \cite{Pinzul:2021cjz}, and massive scalar fields in \cite{Stern:2022}.   These exact solutions will therefore lead to exact expressions for  the corresponding  two-point function on the  boundary. This was shown  in \cite{Pinzul:2021cjz} for the case of massless fields. It is the purpose of this article to obtain the   boundary  two-point function resulting from the exact solutions to the massive field equation found in \cite{Stern:2022}.

The generating function $S|_{\rm on-shell}$ is expressed in terms of the boundary values $\phi_0$ of the solutions for the bulk fields, and the  prescription for computing the  $n-$point connected correlation function for operators ${\cal O}$ located at non-coincidental points $x_i$ on the boundary  is given by
\be\label{adscft}
<{\cal O}(x_1)\cdots{\cal O}(x_n)>=\frac{\delta^n S|_{\rm on-shell}}{\delta \phi_0(x_1)\cdots\delta \phi_0(x_n)}\bigg|_{\phi_0=0} \,.\ee
 For the case of a 2-dimensional  bulk theory, both ${\cal O}$ and  $\phi_0$ are  functions of only one coordinate, the time $t$.  Conformal invariance  severely restricts the form of the $n-$point correlators; For $n=2$ one has
\be
<{\cal O}(t){\cal O}(t ')>\;=\;\,C_{\Delta_+}\,\frac{1 }{|t-t'|^{2\Delta_+}}\,,\label{frmcmttvrslt}
\ee
where $\Delta_+$ is the conformal dimension and $C_{\Delta_+}$ is a constant which can be computed using (\ref{adscft}).  We shall compute the two-point function that results from massive scalar fields on $ncAdS_2$ and show that it has the form given in (\ref{frmcmttvrslt}).  The result for the overall factor $C_{\Delta_+}$  differs from that of the commutative theory.  By taking various limits we can compare with the previous results.

The outline for the remainder of this article is as follows: We first review the calculations for $C_{\Delta_+}$ in the commutative case in section 2, and then apply the analogous procedure to find the constant factor in the non-commutative case in section 3.  Some concluding remarks are made in section 4.

\section{Commutative case}

The analysis of the correspondence principle is more conveniently performed for the Euclidean version of $AdS$, or $EAdS$, due to the absence of propagating states.\footnote{For treatments of  the correspondence principle relying on Lorentzian signature see \cite{Balasubramanian:1998sn,Balasubramanian:1998de,Marolf:2004fy}.} Here we specialize to  $EAdS_2$, closely following the work in \cite{Freedman:1998tz,Minces:1999eg}.

The field equation for a scalar field $\Phi$ with mass $m$
on $EAdS_2$ is
\be  \Delta \Phi\;=\; {m^2}\Phi\,,
\label{XfyXeqmfycmt}
\ee
with $\Delta$  denoting the appropriate Laplacian.
$EAdS_2$ is conveniently parametrized by Fefferman-Graham coordinates $(z,t)$.   They span the half-plane,  $z>0$,  $-\infty<t<\infty$, with the conformal boundary located at $z\rightarrow 0$.\footnote{Some care is needed to define the boundary limit.  One should first assume that the boundary is located at some $z=\epsilon$, and  then take the limit $\epsilon\rightarrow 0$.}  In terms of these coordinates the metric tensor and  Laplacian are given, respectively, by
\be ds^2=\frac{dz^2+dt^2}{z^2}\;,\ee and
\be \Delta=z^2(\partial_z^2+\partial_t^2)\;.\label{cmtvLpl}\ee
The field equation (\ref{XfyXeqmfycmt}) results from the standard scalar field action
\beqa  S&=&\;\;\,\frac 12\int_{R\times R_+} dtdz\left((\partial_z\Phi)^2+(\partial_t\Phi)^2+\frac {m^2}{z^2} \Phi^2\right)\cr
&=&-\frac 12\int_{R\times R_+} \frac{dtdz}{z^2} \,\Phi(\Delta- {m^2})\Phi-\frac 12\int_{R}dt\,\Phi\partial_z\Phi\Big|_{z\rightarrow 0}\,.\label{clsclactn}
\eeqa
Only the boundary term survives when evaluating the action on-shell.

General solutions to (\ref{XfyXeqmfycmt}) that are well behaved away from the conformal  boundary have the form
\be \Phi(z,t)=\frac1{2\pi}\int d\omega\, e^{i\omega  t} \sqrt{ z}\, a(\omega)\, K_{\nu}\left({|\omega|  z}\right)\;,\qquad \nu= \sqrt{
m^2+\frac{1}{4}}\,,\label{gnrlsln}\ee   where  $K_{\nu }(x)$ denotes the modified Bessel function.
To determine the behavior of the solutions near the  conformal boundary we can use\cite{Bateman:100234}
\be K_{\nu }(x)\rightarrow \frac 12 \Bigl(2^{\nu } \Gamma (\nu ) x^{-\nu }+2^{-\nu } \Gamma (-\nu )x^{\nu }\Bigr)\left(1+{\cal O}(x^2)\right),\quad{\rm  as}\;\; x\rightarrow 0\,.\label{ldngKnu}\ee
Then for $0<\nu<1$ (corresponding to  $-\frac 14<m^2<\frac 34 $),  $\Phi(z,t)$ tends towards
\be \frac1{4\pi}\int d\omega\, e^{i\omega  t}  a(\omega)  \left(\left(\frac 2{|\omega|}\right)^{\nu } \Gamma (\nu )  z^{\Delta_- } +\left(\frac {|\omega|}2\right)^{\nu } \Gamma (-\nu ) z^{\Delta_+ }\right),\,{\rm as}\, z=\epsilon\rightarrow 0, \ee
where $\Delta_\pm=\frac 12 \pm \nu$, with $\Delta_+$ to be identified with the conformal dimension.  So upon keeping just the leading term,
 \be \Phi(\epsilon,t)\rightarrow \phi_\epsilon(t)=\epsilon^{\Delta_-}\phi_0(t) \;,\quad{\rm as}\;\;{\epsilon\rightarrow 0}
\,.\label{bndryfy}\;\ee

Here we shall assume Dirichlet boundary conditions.
  $\phi_0(t)$  is regarded as the source field on the boundary, which determines the solution for $\Phi$ in the bulk.\footnote{ For other boundary conditions, see for example \cite{Minces:1999eg}.}  This can be made explicit by writing  the coefficients $a(\omega)$ of the solution (\ref{gnrlsln})  in terms of the Fourier coefficients  $\phi(\omega)$ of the boundary field $ \phi_\epsilon(t)=\frac1{2\pi}\int d\omega e^{i\omega  t}\phi(\omega)$,
\be  \phi(\omega)=\sqrt{ \epsilon} a(\omega) K_{\nu}\left({|\omega| \epsilon}\right)\,.\ee
The solution for  $\Phi(z,t)$ in the bulk can thereby be expressed in terms of these Fourier coefficients
 \be \Phi(z,t)=\frac1{2\pi}\int d\omega e^{i\omega  t}  \phi(\omega)\frac{\sqrt{ z} K_{\nu}\left({|\omega|  z}\right)} {\sqrt{ \epsilon}K_{\nu}\left({|\omega| \epsilon}\right) }\;.\qquad\label{Phindrv} \ee
The evaluation of the on-shell action requires an analogous expression for  $ \partial_z\Phi(z,t)$. For this we can use the identity $\frac d{dx} K_{\nu }(x)=\frac \nu xK_{\nu }(x)-  K_{\nu+1 }(x)$ to write
$$\sqrt z\frac d{dz}\left(\sqrt zK_{\nu }(|\omega|z)\right)= \left(\frac 1{2}+ \nu{} \right)K_{\nu }(|\omega|z)-|\omega| z  K_{\nu+1 }(|\omega|z)\,,$$
and hence
\be \partial_z\Phi(z,t)=\frac1{2\pi}\int d\omega e^{i\omega  t}  \phi(\omega)\frac{\left(\frac 1{2}+ \nu{} \right)K_{\nu }(|\omega|z)-|\omega| z  K_{\nu+1 }(|\omega|z)} {\sqrt{ z\epsilon}K_{\nu}\left({|\omega| \epsilon}\right) }\,.\qquad\label{dzPhi} \ee
Substitution of  (\ref{Phindrv})  and (\ref{dzPhi}) into the boundary term in (\ref{clsclactn}) gives the following result for the on-shell action $ S|_{\rm on-shell}$:
\be-\frac 1{4\pi}\int dt \int dt' \int d\omega e^{i\omega (  t- t')}\phi_\epsilon (t)\phi_\epsilon (t')\,\frac 1\epsilon\left({ \frac 1{2}+ \nu -|\omega|\epsilon \frac{  K_{\nu+1 }(|\omega|\epsilon)}{K_{\nu}\left({|\omega| \epsilon}\right)}} { }\right)\,\Big|_{\epsilon\rightarrow 0}\,.\ee
From (\ref{ldngKnu}), the asymptotic limit of the ratio of Bessel functions is
$$\frac{xK_{\nu+1 }(x)}{K_{\nu }(x)}\rightarrow\frac{ \left(2^{\nu+1 } \Gamma (\nu+1 )+2^{-\nu-1 } \Gamma (-\nu -1)x^{2\nu+2 }\right)}{ \left(2^{\nu } \Gamma (\nu )+2^{-\nu } \Gamma (-\nu )x^{2\nu }\right)}\,\quad{\rm  as}\quad x\rightarrow 0\,,$$
which to leading order is $2\nu\left(1 -\frac{ \Gamma (-\nu )}{\Gamma (\nu )}\left(\frac x2\right)^{2\nu }\right)$ for $ 0<\nu<1$.  Upon applying this result, along with the integral\cite{Minces:1999eg}\be \frac 1{2\pi}\int d\omega e^{-i\omega t}|\omega|^\rho=\frac{2^\rho\Gamma(\frac{\rho+1}2)}{\sqrt{\pi}\Gamma(-\frac\rho 2)}\,\frac 1{|t|^{\rho+1}}\;,\qquad\rho\ne-1,-3,...\;,\label{MinRiv}\ee
one gets\footnote{The singular term
$$ -\frac 1{4\pi} ( \frac 1{2}- \nu)\int dt \int dt' \int d\omega e^{i\omega (  t- t')}\frac{\phi_0 (t)\phi_0 (t')}{\epsilon^{2\nu}}\Big|_{\epsilon\rightarrow 0}\;$$ can be ignored since the integration in $\omega$ leads to a delta function and we are interested in $t\ne t'$.}
\beqa S|_{\rm on-shell}&=&-\frac{ \nu  \Gamma (-\nu )}{2^{2\nu +1}\pi\Gamma (\nu )}\int dt \int dt' \phi_\epsilon (t)\phi_\epsilon (t')\,\epsilon^{2\nu -1}\int d\omega e^{i\omega (  t- t')}{|\omega|^{2\nu } }\,\Big|_{\epsilon\rightarrow 0}\cr&&\cr
&=&-\frac{ \nu }{\sqrt{\pi}}\frac{\Gamma(\nu+\frac{1}2)}{\Gamma (\nu )}\int dt \int dt' \phi_\epsilon (t)\phi_\epsilon (t')\,\epsilon^{2\nu -1}\,\frac 1{|t-t'|^{2\nu+1}}\,\Big|_{\epsilon\rightarrow 0}\,.\label{onshlactn}\eeqa
In terms of the $\epsilon$-independent source field $\phi_0(t)$ in (\ref{bndryfy}) the result is
\beqa S|_{\rm on-shell}&=&-\frac{\nu}{\sqrt{\pi}}\frac{\Gamma(\nu+\frac 12)}{\Gamma({\nu} )}\int dt \int dt' \,\frac {\phi_0 (t)\phi_0 (t')}{|t-t'|^{2\nu+1}}
\,\, .\;\label{clsclrslt}\eeqa
The two-point function for the massive scalar  (\ref{frmcmttvrslt})  is now easily recovered from  (\ref{adscft}) with the resulting factor $C_{\Delta_+}$ given by
\be
C_{\Delta_+}\;=\;\,-\frac{2\nu}{\sqrt{\pi}}\frac{\Gamma(\nu+\frac 12)}{\Gamma({\nu} )}\;.\label{cmttvrslt}
\ee

\section{Non-commutative case}

We now repeat the procedure for the non-commutative case.  The  field  equation  for a massive  scalar field $\hat\Phi$ on non-commutative $EAdS_2$ is
\be  \hat \Delta\hat \Phi
 \;=\; {m^2}\hat \Phi
\label{XfyXeqmfy} \;,\ee where $\hat \Delta$ is the noncommutative version of the Laplacian (\ref{cmtvLpl}). As  in  \cite{Stern:2022}, one can express $\hat \Delta$ in terms of operators $\hat X^{a},\;a=1,2,3,$ which satisfy the $su(1,1)$ algebra
\be [\hat X^a,\hat X^b]=i\alpha\epsilon^{abc} \hat X_c\;,\ee
and have $\hat X^a\hat X_a=-1$. The indices are raised and lowered with the metric diag$(-1,1,1)$, $\epsilon^{abc}$ is totally antisymmetric, with $\epsilon^{012}=1$,  and $\alpha$ is the noncommutative parameter, analogous to Planck's constant, which should correspond to the quantum gravity scale.
The  Laplacian is given by
\be   \frac{\alpha^2}2\hat \Delta\hat \Phi=
 \hat X_{a}\hat \Phi\hat X^{a}\,+\,\hat\Phi\,.
\ee
The field equation (\ref{XfyXeqmfy}) results from the  variation of $\hat\Phi$  in the following action
\be
\hat S=-\frac 1{2\alpha^2} {\rm Tr}\,\Bigl\{[\hat X^a,\hat \Phi] [{\hat X}_{a},\hat \Phi]-(\alpha m)^2\hat  \Phi^2\Bigr\} \label{ncactn}\,.
\ee
The trace in (\ref{ncactn}) can be understood as an integration over symbols of operators, and the integration domain can again be taken to be the half-plane spanned by commuting coordinates $z$ and $t$. As in (\ref{clsclactn}) the action can be spit up into two terms, one of which vanishes on-shell and the other is a boundary term.  Moreover, as was shown  in \cite{Pinzul:2017wch}, the boundary term for the noncommutative theory is {\it identical} to that of the commutative theory.  (A conformal boundary limit can be defined for noncommutative $EAdS_2$ in terms of the representation theory for the $su(1,1)$ algebra.\cite{Pinzul:2017wch}) Thus the on-shell action once again  takes the form
\beqa S|_{\rm on-shell}&=&-\frac 12\int_{R}dt\,\hat \Phi\partial_z\hat \Phi\Big|_{z\rightarrow 0}\,,\label{ncclsclactn}
\eeqa
where here $ \hat \Phi$ actually denotes the symbol of  the field.

As shown in \cite{Stern:2022},  (\ref{XfyXeqmfy}) has exact solutions, which are given in terms of generalized Legendre functions.  Upon restricting to fields that are well-behaved in the bulk, and integrating over all frequencies $\omega$, one has
\be\hat \Phi =\frac1{2\pi}\int d{\omega}\, a(\omega)  e^{i\omega\hat t/2}\,P_{-\Delta_- }^{-\frac {2\kappa}\alpha}\left(\frac{2\hat r }{{|\omega}|  \alpha}\right)\,e^{i\omega\hat t/2}\;,\label{ncslnfy}\ee where   $\hat t$ and $\hat r$ are noncommutative operators, which can be used to construct $\hat X^a$ (see \cite{Stern:2022}), and $\kappa=\sqrt{1+\frac{\alpha^2}4}$.    $\hat t$ and $\hat r$  satisfy the commutator  $ [\hat t,\hat r]=-i\alpha\,.$

As in the commutative case, the  solution can be expressed in terms of boundary fields. The symbol of $\hat r^{-1}$ tends to zero in this limit, and the algebra   of functions of $\hat r$ and $\hat t$ is effectively commutative near the boundary.   Therefore the symbol of the solution (\ref{ncslnfy}) has the  conformal boundary limit:
\be\phi_\epsilon(t)=\frac1{2\pi}\int d\omega\, e^{i\omega  t}  a(\omega) \,P_{-\Delta_- }^{-\frac{ 2\kappa}\alpha}\left(\frac{2}{{|\omega}|  \alpha\epsilon }\right)\,, \;\;\;{\rm as}\; \epsilon\rightarrow 0\,.\ee
From \cite{Stern:2022} this tends to
 $\epsilon^{\Delta_-}\phi_0(t)\; {\rm as}\;{\epsilon\rightarrow 0}$ for $\nu>0$, just as in the commutative case (\ref{bndryfy}).
We shall again assume Dirichlet boundary conditions with  $\phi_0(t)$  regarded as the source field.  Re-expressing the solution (\ref{ncslnfy}) in terms  of the Fourier transform $\phi(\omega) $ of the boundary field $ \phi_\epsilon(t)$,
\be \phi(\omega)   =a(\omega) \,P_{-\Delta_- }^{-\frac {2\kappa}\alpha}\left(\frac{2}{{|\omega}|  \alpha\epsilon }\right)\;,\ee
we get
\be\hat \Phi =\frac1{2\pi}\int d\omega\, \phi(\omega)  e^{i\omega\hat t/2}\,\frac{P_{-\Delta_- }^{-\frac {2\kappa}\alpha}\left(\frac{2\hat r }{{|\omega}|  \alpha}\right)}{P_{-\Delta_- }^{-\frac {2\kappa}\alpha}\left(\frac{2}{{|\omega}|  \alpha\epsilon }\right)}\,e^{i{|\omega}|\hat t/2}\,.\ee

Once again we need to compute the derivative of the solution with respect to the radial coordinate. This, in general, will introduce operator ordering ambiguities. However, such ambiguities are not a concern  for the computation of the on-shell action, since we  only need the result in the  conformal boundary limit where the coordinates effectively commute. So let us choose the radial derivative to be $\frac i\alpha[\hat t,\hat\Phi]$.  Then the  symbol of $\frac i\alpha[\hat t,\hat\Phi]$ in the boundary limit is  $-z^2\partial_z\hat \Phi|_{z\rightarrow 0}$.
After applying the identity\cite{gradshteyn2007}
  \be\frac {dP^\mu_\rho(x)}{dx}=\frac 1{1-x^2}\left({(\rho +1) x P_{\rho }^{\mu }(x)-(-\mu +\rho +1) P_{\rho +1}^{\mu }(x)}\right)\;,\ee
we get the following result for $\partial_z\hat \Phi|_{z=\epsilon}$ as $\epsilon\rightarrow 0$
$$-\frac1{4\pi}\int d{\omega} e^{i{\omega}  t}  \phi({\omega})  {\alpha  {|\omega}| }\left((\Delta_- -1) \Bigl(\frac{2}{ |\omega|  \alpha  \epsilon
 }\Bigr)+\Bigr(\Delta_++\frac {2\kappa}\alpha\Bigl)\frac{P^{-\frac {2\kappa}\alpha}_{{\Delta_+ }}\left(\frac{2}{ {|\omega}|
   \alpha \epsilon }\right)}{
P_{-\Delta_- }^{-\frac {2\kappa}\alpha}\left(\frac{2}{{|\omega}| \alpha \epsilon }\right)}\right)\,.$$
 The substitution of this result in the on-shell action
(\ref{ncclsclactn})  yields
\beqa &&S|_{\rm on-shell}=\frac 1{8\pi}\int dt \int dt' \int d{\omega} e^{i{\omega} (  t- t')}\phi_\epsilon (t)\phi_\epsilon (t')\;\times\cr&&\cr&&\; {\alpha  {|\omega}| }\left((\Delta_- -1) \left(\frac{2}{ {|\omega}|
   \alpha  \epsilon}\right)+\left(\Delta_++\frac{ 2\kappa}\alpha\right)\frac{P^{-\frac 2\alpha\kappa}_{{\Delta_+ }}\left(\frac{2}{ {|\omega}|
   \alpha\epsilon }\right)}{
P_{-\Delta_- }^{-\frac 2\alpha\kappa}\left(\frac{2}{{|\omega}|  \alpha\epsilon }\right)}\right)\Bigg|_{\epsilon\rightarrow 0}.\label{osais}\eeqa
As with the commutative answer, the term that yields a  delta function after integration in $\omega$ can be dropped.  To evaluate the remaining term we need the asymptotic behavior of ${yP^\mu_{\nu+\frac 12}(1/y)}\,/\,{ P^\mu_{\nu-\frac 12}(1/y)}$
as $y\rightarrow 0$, which is
$$\frac{2\nu}{\nu-\mu+\frac12}\left(1-\frac{\Gamma(-\nu)\Gamma(\nu-\mu+\frac12)}{2^{2\nu}\Gamma(\nu)\Gamma(-\nu-\mu+\frac12)}y^{2\nu}\right)\;,
\quad{\rm for}\;\;0<\nu<1\,.$$
Applying this to (\ref{osais}) gives
$$S|_{\rm on-shell}=\frac 1{4\pi}\int dt \int dt' \int d{\omega} e^{i{\omega} (  t- t')}\phi_\epsilon (t)\phi_\epsilon (t')\frac{1}{\epsilon
  }\, \times\,\qquad\qquad$$
\be \frac{2\nu(\Delta_++\frac {2\kappa}\alpha)}{\nu+\frac{ 2\kappa}\alpha+\frac12}\Biggr(1-\frac{\Gamma(-\nu)\Gamma(\nu+\frac {2\kappa}\alpha+\frac12)}{2^{2\nu}\Gamma(\nu)\Gamma(-\nu+\frac {2\kappa}\alpha+\frac12)}\Bigl(\frac {{|\omega}|\alpha \epsilon}2\Bigr)^{2\nu}\Biggr)\Bigg|_{\epsilon\rightarrow 0}\,.\ee
The integral over $\omega$ can be performed, again using
 (\ref{MinRiv}), with the  result for the  on-shell action being
$$-\frac{\nu}{\sqrt{\pi}}\frac{\Gamma(\nu+\frac{1}2)\Gamma(\nu+{\frac {2\kappa}\alpha}+\frac12)}{\Gamma(\nu)\Gamma(-\nu+{\frac {2\kappa}\alpha}+\frac12)}\int dt \int dt' \phi_\epsilon (t)\phi_\epsilon (t')\frac{1}{\epsilon
  }\Bigl(\frac {\alpha \epsilon}2\Bigr)^{2\nu}\frac 1{|t-t'|^{2\nu+1}}\Bigg|_{\epsilon\rightarrow 0}\,.$$
In terms of the $\epsilon -$independent source fields $\phi_0$ this becomes
$$ S|_{\rm on-shell}=-\left(\frac {\alpha }2\right)^{2\nu}\frac{\nu}{\sqrt{\pi}}\frac{\Gamma(\nu+\frac{1}2)\Gamma(\nu+{\frac {2\kappa}\alpha}+\frac12)}{\Gamma(\nu)\Gamma(-\nu+{\frac {2\kappa}\alpha}+\frac12)}\,\int dt \int dt' \,\frac {\phi_0 (t)\phi_0 (t')}{|t-t'|^{2\nu+1}}\,.$$
The conformal answer for the two-point function (\ref{frmcmttvrslt})  is then once again recovered after using  (\ref{adscft}).    Now the  overall  factor is
\be
C^{\rm nc}_{\Delta_+}\;=\;\,-\left(\frac {\alpha }2\right)^{2\nu}\frac{2\nu}{\sqrt{\pi}}\frac{\Gamma(\nu+\frac{1}2)\Gamma(\nu+{\frac {2\kappa}\alpha}+\frac12)}{\Gamma(\nu)\Gamma(-\nu+{\frac {2\kappa}\alpha}+\frac12)}\,.\label{ncCpls}
\ee

The ratio of the noncommutative result to the commutative result (\ref{cmttvrslt}) is given by the simple expression
\be {\cal R}^{(2)}(\alpha,\nu)= \left(\frac {\alpha }2\right)^{2\nu}\frac{\Gamma(\nu+{\frac{ 2\kappa}\alpha}+\frac12)}{\Gamma(-\nu+{\frac {2\kappa}\alpha}+\frac12)}\label{thrto}\,. \ee
One can  check various limits.
The ratio smoothly goes to one in the commutative  limit $\alpha\rightarrow 0$.  It reduces to $\kappa$ in the massless case, $\nu=\frac 12$, which agrees with the result found in \cite{Pinzul:2021cjz}.\footnote{If one does a leading order expansion  in $\alpha^2$ one gets
$$ {\cal R}^{(2)}(\alpha,\nu)=1+\frac\nu{12}\left(\frac{13}4-\nu^2\right)\alpha^2+{\cal O}(\alpha^4).$$
This disagrees with the result found in \cite{deAlmeida:2019awj}, which is due to the fact that  a different regularization scheme was used in that approach.  The answers agree on the other hand, if one re-evaluates integrals  in  \cite{deAlmeida:2019awj} with the regularization scheme used here.  Both regularizations give the same result in the massless case. For a discussion of the different regularizations see the appendix in \cite{Freedman:1998tz}.

We also note that the result found here agrees with that found in \cite{Ydri:2021xpw} for the massless case, but it differs when $\nu\ne \frac 12$.}

\section{Concluding Remarks}
It had previously been conjectured that the boundary two-point function associated with field theory on a noncommutative $AdS_2$ bulk satisfies the constraints of conformal invariance, i.e. it has the form  (\ref{frmcmttvrslt}).  This was previously shown to be true  up to leading order in the noncommutativity parameter $\alpha$. In this article  we proved that the conjecture is true to all orders in $\alpha$, at least for the scalar field. A similar proof should be possible for spinors. The result means that the noncommutativity of the bulk only effects  the over-all normalization of  the boundary two-point function.

It is a nontrivial step to see whether or not these results can be extended to the case of  $n(>2)-$point correlators on the boundary.  This will require analyzing a fully interacting field theory on the noncommutative bulk.  If conformal invariance is satisfied for the $n(>2)-$point boundary correlators then, once again, only the  over-all factor is effected by  the noncommutative bulk. We would then have an expression for the ratio $ {\cal R}^{(n)}(\alpha,\nu)$ of the over-all factor  for the noncommutative correlator with that of the commutative correlator. If the ratio is such that $\sqrt[n]{ {\cal R}^{(n)}(\alpha,\nu)}=\sqrt{ {\cal R}^{(2)}(\alpha,\nu)}$, it would suggest
that the only effect that  bulk noncommutativity has on the boundary is to renormalize the conformal operators ${\cal O}$.  If instead this turns out not to be  the case, we would  conclude that  the  couplings between boundary operators pick up quantum gravity corrections as well.
Only a preliminary investigation in this direction has been undertaken so far.
A  cubic interaction term was introduced to the massive scalar field  on noncommutative $EAdS_2$  in \cite{deAlmeida:2019awj}.  The analysis there was quite nontrivial because only a perturbative solution  (in $\alpha$) was available, and one  needs to perturb in the other  parameter,  the coupling constant $\lambda$, as well. Although there were strong indications that the leading order correction to $3-$point function satisfied the constraints of conformal invariance, we were not able to write down an explicit expression for the leading order correction, in either of the parameters.  The fact that we now  have an exact answer for the free theory, could be quite beneficial  for obtaining the $3-$point function, since then we would only have to perform an expansion in one parameter, i.e., $\lambda$.  The only obstacle to proceeding with the analysis is the construction of a meaningful Green function on the noncommutative bulk.  We hope to report on this construction in a future work.

\end{document}